\documentclass[sigconf,natbib=true,anonymous=false]{acmart}

\usepackage{amssymb}

\usepackage{amsthm,amsmath,amssymb}
\usepackage{mathrsfs}
\usepackage{cleveref}
\usepackage{enumitem}
\usepackage{multirow}
\usepackage{multicol}

\usepackage{graphicx}
\usepackage{booktabs}
\usepackage{array}
\usepackage{tabularx} 
\AtBeginDocument{%
  }

\setcopyright{acmlicensed}
\copyrightyear{2025}
\acmYear{2025}
\acmDOI{10.1145/3726302.3729903}
\acmConference[SIGIR 2025]{The 48th International ACM SIGIR Conference on Research and Development in Information Retrieval}{July 13--18, 2025}{Padua, Italy}
\acmISBN{979-8-4007-1592-1/2025/07}

\begin{document}

\title{Beyond Whole Dialogue Modeling: Contextual Disentanglement for Conversational Recommendation}

\author{Guojia An}
\orcid{1234-5678-9012}
\affiliation{%
  \institution{University of Electronic Science and Technology of China}
  \city{Chengdu}
  \state{Sichuan}
  \country{China}
}
\email{anguojia.1999@gmail.com}

\author{Jie Zou}
\authornote{Corresponding author.}
\affiliation{%
  \institution{University of Electronic Science and Technology of China}
  \city{Chengdu}
  \state{Sichuan}
  \country{China}
}
\email{jie.zou@uestc.edu.cn}

\author{Jiwei Wei}
\affiliation{%
  \institution{University of Electronic Science and Technology of China}
  \city{Chengdu}
  \state{Sichuan}
  \country{China}}
\email{mathematic6@gmail.com}

\author{Chaoning Zhang}
\affiliation{%
  \institution{University of Electronic Science and Technology of China}
  \city{Chengdu}
  \state{Sichuan}
  \country{China}}
\email{chaoningzhang1990@gmail.com}

\author{Fuming Sun}
\affiliation{%
  \institution{Dalian Minzu University}
  \city{Dalian}
  \state{Liaoning}
  \country{China}}
\email{sunfuming@dlnu.edu.cn}

\author{Yang Yang}
\affiliation{%
  \institution{University of Electronic Science and Technology of China}
  \city{Chengdu}
  \state{Sichuan}
  \country{China}}
\email{yang.yang@uestc.edu.cn}
\renewcommand{\shortauthors}{Guojia An et al.}

\begin{abstract}
  Conversational recommender systems aim to provide personalized recommendations by analyzing and utilizing contextual information related to dialogue. However, existing methods typically model the dialogue context as a whole, neglecting the inherent complexity and entanglement within the dialogue. Specifically, a dialogue comprises both focus information and background information, which mutually influence each other. Current methods tend to model these two types of information mixedly, leading to misinterpretation of users’ actual needs, thereby lowering the accuracy of recommendations. To address this issue, this paper proposes a novel model to introduce contextual disentanglement for improving conversational recommender systems, named DisenCRS. The proposed model DisenCRS employs a dual disentanglement framework, including self-supervised contrastive disentanglement and counterfactual inference disentanglement, to effectively distinguish focus information and background information from the dialogue context under unsupervised conditions. Moreover, we design an adaptive prompt learning module to automatically select the most suitable prompt based on the specific dialogue context, fully leveraging the power of large language models. Experimental results on two widely used public datasets demonstrate that DisenCRS significantly outperforms existing conversational recommendation models, achieving superior performance on both item recommendation and response generation tasks.
\end{abstract}





\maketitle

\section{INTRODUCTION}
Traditional recommender systems provide personalized recommendations by mining user behavior. Although they have achieved some success, they still face the limitation in capturing users' real-time intentions \cite{mb16,mb17,mb33}. To address this problem, conversational recommender systems (CRS) have emerged, which can dynamically understand user needs through natural language interactions to provide personalized and effective recommendations. With the rapid advancements of CRS in fields such as e-commerce, movie recommendation, and travel planning, it has become a prominent research direction in both academia and industry.

In early studies, researchers find that users usually express their preferences through specific entities, prompting some studies to focus on modeling entity-related information \cite{mb24,mb25,mb26,mb30}. Although this method may be effective in certain cases, focusing only on specific entities while ignoring the context of the conversation may lead to the loss of information, thereby misunderstanding the user’s true intent. Fortunately, with the advancement of natural language processing technologies, the model's ability to understand the semantics of the conversation context has become increasingly powerful. Therefore, recent studies have shifted focus toward modeling the conversation context as a whole \cite{mb34,mb38,mb35,mb27}. For example, $\text{C}^2\text{-}\text{CRS}$
 \cite{mb36} employs transformers to model the entire conversation context. Then, UniCRS \cite{mb27} and DCRS \cite{mb35} leverage RoBERTa \cite{mb45} to capture the semantic information throughout the dialogue. 

\begin{figure}[t]
	\begin{center}
		\includegraphics[width=0.98\linewidth]{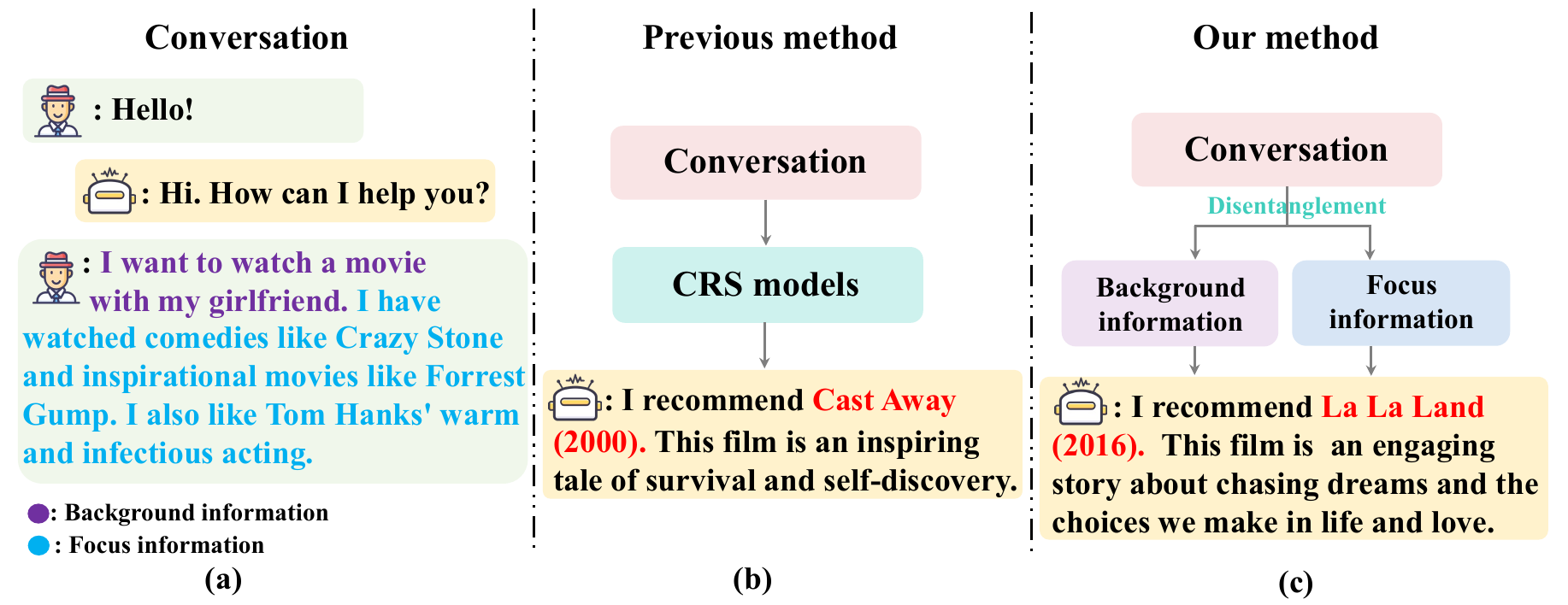}
	\end{center}
	\caption{Previous CRS method vs. our method.}
	\label{fig: f1}
\end{figure}

\begin{figure}[t]
	\begin{center}
		\includegraphics[width=1\linewidth]{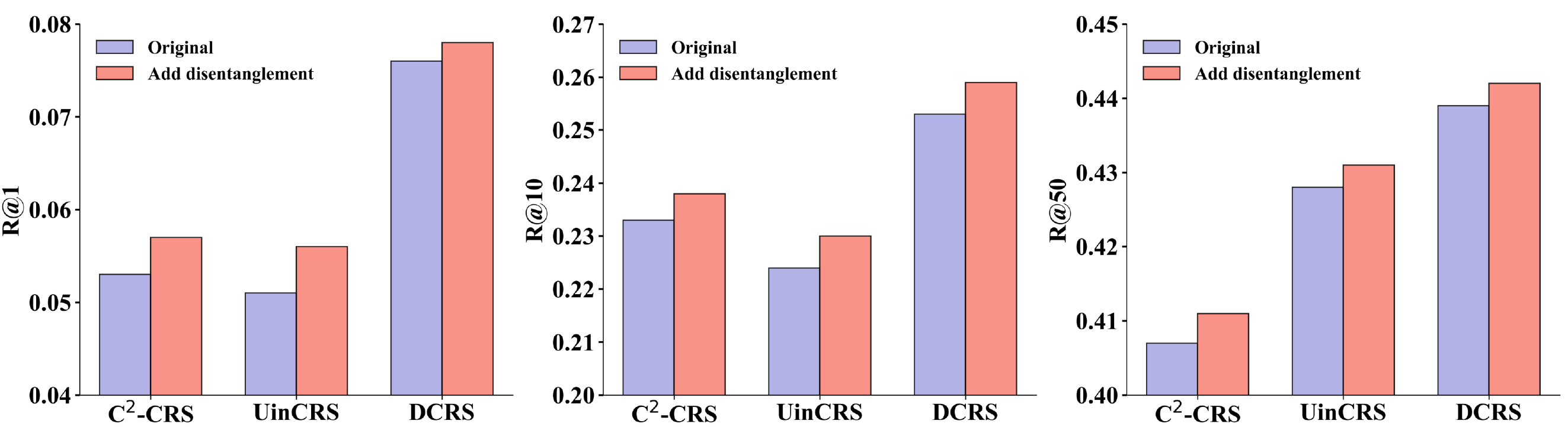}
	\end{center}
	\caption{Performance comparison of the original method and the method with added disentanglement.}
	\label{fig: f2}
\end{figure}

\textbf{Motivation:} 
Although existing methods of modeling the conversation context as a whole have shown promising results in practice, they often neglect the complexity and entanglement of information within dialogue contexts. Specifically, as shown in Figure \ref{fig: f1}(a), the information within a conversation contains focus information (contextual information related to entities) and background information (contextual information unrelated to entities). Previous methods treat the entire dialogue context uniformly and model these two types of information mixedly, which prevents the model from effectively distinguishing the importance of different pieces of information, thereby hindering its ability to accurately capture the user's actual intent. 
For example, in the case illustrated in Figure \ref{fig: f1}(b), the previous methods primarily concentrate on the focus information while ignoring the background information that the user wants to watch a movie with his/her girlfriend, leading to wrongly recommending the movie `Cast Away'. 
In contrast, our approach first separates these two types of information and then dynamically balances them based on the dialogue context. As illustrated in Figure \ref{fig: f1}(c), our approach recommends `La La Land', which is the result of dynamically integrating focus information and background information (i.e., go watch a movie with his/her girlfriend). 
The advantage of this disentangled modeling approach lies in its ability to effectively distinguish the importance of different types of information, reducing interference from irrelevant details \cite{a4,a5}. Also, it flexibly utilizes background information as supplementary input when necessary, thereby enhancing the accuracy of identifying the user's primary intent. Furthermore, from the experimental results, as shown in Figure \ref{fig: f2}, we compare the performance differences between the dialogue disentanglement approach and the existing holistic modeling method across three state-of-the-art conversational recommendation models. The results show that the disentanglement approach can consistently improve performance, further confirming the necessity and effectiveness of disentanglement for conversational context. 

\textbf{Challenges:} Although it is essential to separate the focus information and background information from the conversation context, there are two major challenges to utilize these two types of information: (1) In the absence of supervisory signals, how to effectively disentangle focus information and background information from the conversation context has not been fully explored. (2) In different conversation scenarios, how to utilize dynamic adjustment mechanisms to balance the focus information and background information to achieve accurate identification of the user's true intentions remains challenging. 

\textbf{Our Work:} To tackle the aforementioned challenges, we propose a novel model of Contextual \textbf{Disen}tanglement for \textbf{C}onversational \textbf{R}ecommender \textbf{S}ystem (\textbf{DisenCRS}). Specifically, we design a \textit{dual contextual disentanglement module}, which effectively decomposes complex dialogue context information into focus information and background information under unsupervised conditions. This module comprises two components: (1) contrastive disentanglement, which establishes a proxy mechanism to guide the disentanglement of the context, and (2) counterfactual inference disentanglement, which aims to capture the impact of the remaining factor by eliminating the influence of either the background information or the focus information factor. Further, we design an \textit{adaptive prompt learning module} that can accurately identify and utilize key information in different conversation scenarios through a dynamic adjustment mechanism to address the second challenge mentioned above. In particular, we treat focus and background information as two types of prompts and construct a prompt pool through weighted fusion. A learnable prompt selector is then introduced to dynamically select the most appropriate prompts from the prompt pool according to the specific dialogue context. 
In summary, the main contributions of this paper are as follows:
\begin{itemize}[itemsep=0pt, topsep=0pt]
\item {We emphasize and analyze the importance of disentangling focus and background information from complex conversation contexts. To the best of our knowledge, this paper is the first effort to introduce contextual disentanglement into conversational recommendation, providing a new perspective for subsequent research on conversational recommendation.}
\item {We propose a new CRS model, called DisenCRS, which is equipped with contextual disentanglement and adaptive prompt learning, to effectively take advantage of focus information and background information for improving conversational recommendation. 
}
\item {Extensive experiment results on two benchmark datasets demonstrate that the proposed model outperforms state-of-the-art methods, achieving statistically significant performance improvements in both recommendation and response generation tasks.}
\end{itemize}

\begin{figure*}[t]
	\begin{center}
		\includegraphics[width=1\linewidth]{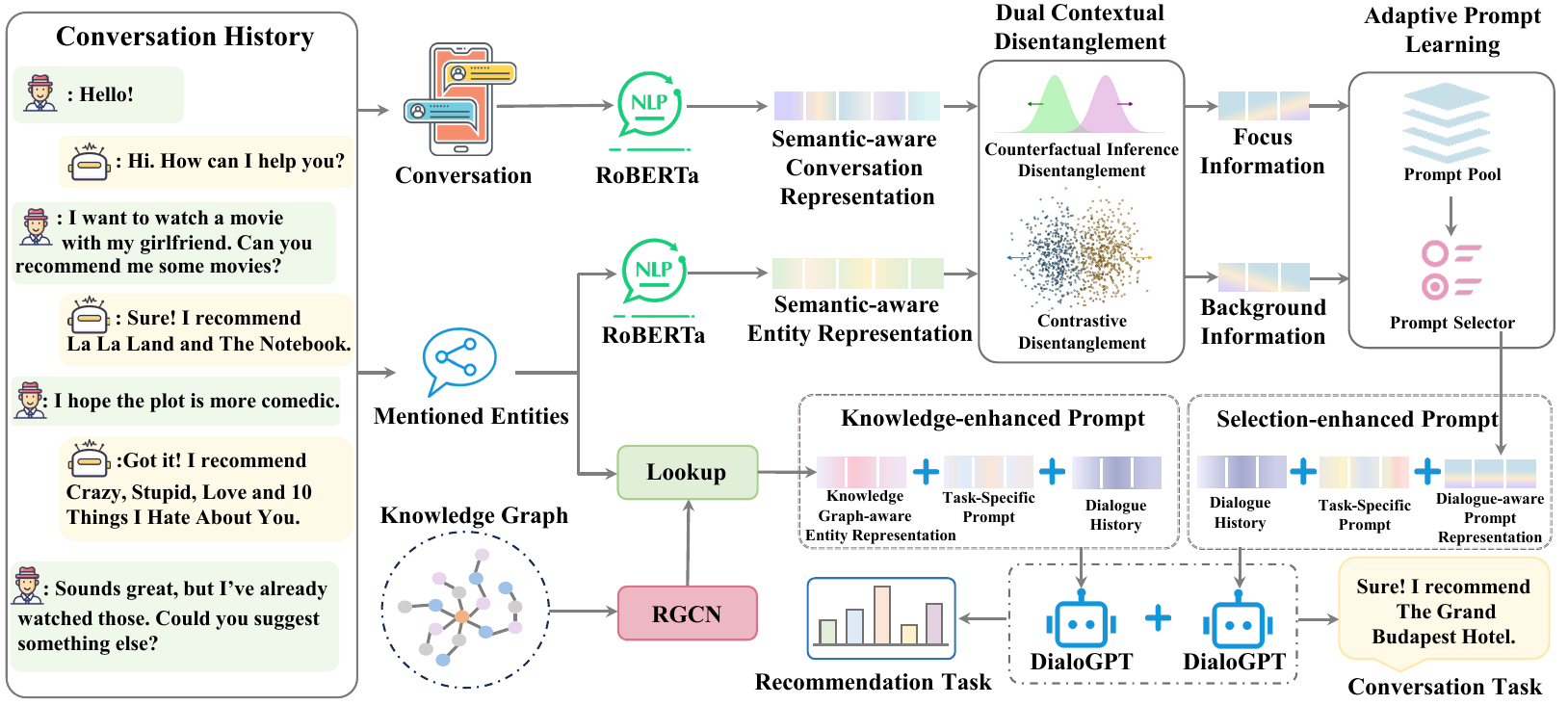}
	\end{center}
	\caption{The overall architecture of DisenCRS, which consists of four main components: (1) the representation learning module, which learns the representations essential for our task; (2) the dual contextual disentanglement module, which separates focus information and background information from the conversation context; (3) the adaptive prompt learning module, which dynamically adjusts prompts based on the input dialogue context; (4) the downstream task adaption module, which utilizes the outputs from the previous modules to perform specific tasks.}
	\label{fig: f3}
\end{figure*}

\section{RELATED WORK}
\subsection{Conversational Recommender Systems}
Existing CRS research can be categorized into two major directions: attribute-based CRSs \cite{mb01,mb03,mb06,mb09,mb11} and generation-based CRSs \cite{mb13,mb26,mb27,mb34,a3}. The former focuses on accurately capturing user preferences in as few interaction rounds as possible by asking the user clarifying questions \cite{a1,a2} about item attributes. In contrast, the latter not only emphasizes the accuracy of recommendation results but also highlights the importance of the fluency and quality of the generated natural language responses.

Attribute-based conversational recommender methods rely on predefined templates to interact with users (i.e., asking users about their preferences for item attributes or making recommendations) \cite{mb04,mb10,mb14}. Such CRSs aim to achieve high accuracy of recommendations while with a focus on designing effective strategies to accurately understand user intentions within the minimum number of interaction rounds. They usually adopt reinforcement learning \cite{mb02,mb05,mb07,mb08} or bandit-based solutions \cite{mb12} in such systems to perform decision-making for whether to ask a question or recommend items. By leveraging such decision-making methods, CRSs can continuously adjust recommendation strategies based on user feedback to maximize long-term returns.

Generation-based CRSs have attracted widespread attention due to their ability to generate more natural and fluent dialogue content. 
In early research of generation-based CRSs, many scholars recognized that entity information often effectively represents user intentions and plays a crucial role in improving the accuracy of recommendations. For instance, \citet{mb24} firstly introduce knowledge graphs to CRSs to model entity information in dialogues. Building on this, KGSF \cite{mb26} incorporate both word-oriented and entity-oriented knowledge graphs to enhance data representations in CRSs. Subsequently, \citet{mb32} further models the sequential relationships between entities to better utilize entity information. \citet{mb15} attempt to enrich the semantic representations of items by leveraging user reviews of the items. With the remarkable advancements in natural language understanding technologies, leveraging complete dialogue context information has become a popular trend for capturing users' complex interest preferences. Specifically, $\text{C}^2$-CRS \cite{mb36} employs a Transformer architecture to model dialogue context. They integrate entity information and review data to generate the final user representation through a coarse-to-fine contrastive learning approach. Then, \citet{mb27} utilize the pre-trained language model RoBERTa to represent dialogue context, emphasizing the modeling of consistency between dialogue tasks and recommendation tasks, thereby enhancing the task synergy performance in CRS. On this basis, DCRS \cite{mb35} also utilizes RoBERTa to model dialogue context and further develops a demonstration-augmented prompt learning approach for conversational recommendation. 

Our work belongs to the field of generation-based CRSs, but has significant differences compared to existing approaches. Although we also utilize conversational context information, we place special emphasis on the complexity and entanglement of context information. Instead of directly using the entire conversation context, we propose a novel method that explicitly distinguishes between focus and background information in the dialogue context and adaptively balances the use of these two types of information, thereby enhancing both recommendation accuracy and response quality.

\subsection{Disentanglement for Recommendation}
Disentanglement in recommendation refers to effectively separating different features and factors from the latent embedding space, enabling a clearer understanding and representation of user interests \cite{mb23}. This approach enhances both the accuracy and interpretability of recommendation systems \cite{mb47}. Recent disentanglement methods aimed to differentiate various reasons in the context of recommendation, such as user interest and consistency \cite{mb18}, long-term and short-term interests \cite{mb19}, user intent \cite{mb21,mb22}, ID and modality \cite{mb20}. Unlike the aforementioned disentangled recommendation methods, this study attempts to separate the entangled information inherent in dialogue context within the conversational recommendation scenario. 
 
\section{METHODOLOGY}  
In this section, we first define the task and provide a detailed introduction to our proposed DisenCRS model, the architecture of which is shown in Figure \ref{fig: f3}.

\subsection{Problem Formalization}
Given a dialogue $\mathcal{C}$ in CRS, it is composed of a sequence of utterances, which can be expressed as $\mathcal{C} =\left\{\mathcal{U}_{i}\right\}_{i=1}^{n_u}$ ($n_u$ represents the number of utterances in the dialogue). And, $\mathcal{U}_{i}=\left\{\mathcal{W}_{j}\right\}_{j=1}^{n_w}$ ($n_w$ represents the number of words in the utterance) represents a sequence of words from the vocabulary $\mathcal{V}$ in the \textit{j}-th utterance. 

Formally, the tasks in CRS can be expressed as follows: For the recommendation task, given a dialogue $\mathcal{C}$, the goal is to select an appropriate subset of items $\mathcal{I}_{\text{rec}}$ that align with the user's intent from the candidate item set $\mathcal{I}$ ($\mathcal{I}_{\text{rec}} \subseteq \mathcal{I}$), and then recommend them to the user. For the response generation task, the goal is to generate an accurate response $\mathcal{R} =\left\{\mathcal{U}_{i}\right\}$ based on the dialogue $\mathcal{C}$.

\subsection{Representation Learning Module}
In this section, we present our representation learning module, which encodes three types of representations: knowledge graph-aware entity representation, semantic-aware entity representation, and semantic-aware conversation representation.

\subsubsection{Knowledge Graph-aware Entity Representation} To accurately model the representation of entities in the latent space, we follow existing research \cite{mb26,mb27} and adopt a knowledge graph-based approach. Specifically, we construct a knowledge graph $\mathcal{G}$ using an entity set $\mathcal{N}$ and a relation set $\mathcal{R}$, where semantic facts are represented in the form of triples $\langle e_1, r, e_2 \rangle$, with \(e_1, e_2 \in \mathcal{E}\), and \(r \in R\). Subsequently, we employ a relational graph convolutional network (RGCN) \cite{mb44} to encode the knowledge graph $\mathcal{G}$. Formally, this process can be expressed as:
\begin{equation}
\mathbf{h}_{e}^{(l+1)}=\sigma\left(\sum_{r \in \mathcal{R}} \sum_{e^{\prime} \in \mathcal{E}_{e}^{r}} \frac{1}{Z_{e, r}} \mathbf{W}_{r}^{(l)} \mathbf{h}_{e^{\prime}}^{(l)}+\mathbf{W}_{e}^{(l)} \mathbf{h}_{e}^{(l)}\right),
\end{equation}
where $\mathbf{h}_{e}^{(l+1)}\in \mathbb{R}^{d_e}$ represents the embedding of entity $e$ in the latent space at layer $l+1$. ${d_e}$ defines the dimensions of entity representations. $\sigma$ is the ReLU activation function, which enhances the non-linear expressiveness of the feature representation. $\mathcal{E}_{e}^{r}$ denotes the set of all neighboring nodes connected to node $e$ through relation $r$. $\mathbf{W}_{r}^{(l)}$ and $\mathbf{W}_{e}^{(l)}$ are learnable weight matrices. ${Z_{e, r}}$ is the normalization factor, representing the number of neighbors. 

As the dialogue context may involve multiple entities, we plan to use a self-attention mechanism to combine all entity representations, thereby obtaining a unified knowledge graph-aware conversation representation $\mathbf{h}_\mathcal{G}$. The formula is shown as follows:
\begin{equation}
\label{eq:2}
\boldsymbol{\alpha}_1 = \operatorname{softmax}\left(\mathbf{b}_1^{\mathrm{T}} \tanh \left(\mathbf{W}_{a} \mathbf{H}_e^g\right)\right),
\end{equation}
\begin{equation}
\label{eq:3}
    \mathbf{h}_{\mathcal{G}} = \mathbf{H}_e^g \boldsymbol{\alpha}_1^{\mathrm{T}},
\end{equation}
where $\boldsymbol{\alpha}_1$ represents the weights of all mentioned entities in the context, $\mathbf{H}_e^g \in \mathbb{R}^{d_{e} \times n_{e}}$ is the entity embedding matrix, $n_e$ is the number of entities mentioned in the context $\mathcal{C}$, $\mathbf{W}_{a}\in \mathbb{R}^{d_{e} \times d_{e}}$, $\mathbf{b}_1\in \mathbb{R}^{d_{e}}$ are learnable transformation matrix.

\subsubsection{Semantic-aware Entity Representation}
In addition to modeling entities through the knowledge graph, we also use RoBERTa to enhance semantic-aware embeddings of entities. Specifically, we extract an entity set from the dialogue context and encode them with RoBERTa to generate semantic-aware entity representation $\mathbf{H}_e^s \in \mathbb{R}^{d_{e} \times n_{e}}$.
\begin{equation}
\mathbf{H}_e^s = \text{RoBERTa}(\mathcal{N}_{\mathcal{C}_k}), 
\end{equation}
where \text{RoBERTa} \cite{mb45} is a pre-trained language model, $\mathcal{N}_{\mathcal{C}_k}$ denotes the set of entities in conversation $\mathcal{C}_k$. Subsequently, similar to \cref{eq:2,eq:3}, a comprehensive semantic representation of entities $\mathbf{h}_p$ is obtained.

\subsubsection{Semantic-aware Conversation Representation} In addition to entity mentions, a conversation $\mathcal{C}_{k}$ also encompasses other rich contextual features that are essential for accurately capturing and representing user interests. Therefore, we also utilize RoBERTa to encode the dialogue context, generating each token representation $\mathbf{h}_c$. The formulation is as follows:
\begin{equation}
\mathbf{h}_c = \text{RoBERTa}(\mathcal{C}_{k}).
\end{equation}
Additionally, to obtain the global semantic representation of the entire dialogue context, we use the embedding vector of the [CLS] token in RoBERTa. This allows RoBERTa to effectively model the complex semantic relationships within the context and produce semantic-aware conversation representation $\mathbf{h}_{cls}$.

\subsection{Dual Contextual Disentanglement Module}
In this section, we propose the dual contextual disentanglement module, which includes two self-supervised approaches, i.e., contrastive disentanglement and counterfactual inference disentanglement, to disentangle focus information and background information within the dialogue context. 
\subsubsection{Contrastive Disentanglement}
We propose a contrastive disentanglement strategy to ensure that the embedding vectors of focus information and background information are significantly distinguishable in the representation space. Since explicit labels for focus and background information are not provided in the dialogue context, this presents a challenge for disentangling them. Therefore, we extract entity-related information from the dialogue context as proxy signals to guide the disentanglement of focus and background information. Specifically, we design a contrastive learning task where the similarity between focus information and the proxy information is significantly higher than the similarity between background information and the proxy information. similarly, the similarity between focus information and the proxy information should also be significantly higher than the similarity between focus information and background information. The contrastive process is formulated as follows:
\begin{equation}
\operatorname{sim}\left(\boldsymbol{\mathbf{h}}_f, \boldsymbol{\mathbf{h}}_p\right)>\operatorname{sim}\left({\mathbf{h}}_f, \boldsymbol{\mathbf{h}}_b\right), 
\label{eq:xsd}
\end{equation}
\begin{equation}
    \operatorname{sim}\left(\boldsymbol{\mathbf{h}}_f, \boldsymbol{\mathbf{h}}_p\right)>\operatorname{sim}\left({\mathbf{h}}_b, \boldsymbol{\mathbf{h}}_p\right),
\label{eq:xsd2}
\end{equation}
where $\mathbf{h}_f$ and $\mathbf{h}_b$  represent the focus information and background information, which are disentangled from the dialogue semantic representation $\mathbf{h}_{cls}$. We use semantic-aware entity representation $\mathbf{h}_p$ as the proxy information and use cosine similarity $\operatorname{sim}(x, y)=\frac{x \cdot y}{\|x\| y \|}$ as the similarity function. Finally, we implement loss functions based on triplet loss to accomplish the contrastive process in \cref{eq:xsd,eq:xsd2}. Formally, the loss functions based on Euclidean distance are computed as follows:
\begin{equation}
\begin{split}
    \mathcal{L}_{c d}=&\max \left(0, {\operatorname{sim}}\left(\mathbf{h}_f, \mathbf{h}_b\right)-{\operatorname{sim}}\left(\mathbf{h}_f, \mathbf{h}_p\right)+m\right) + \\
    &\max \left(0, {\operatorname{sim}}\left(\mathbf{h}_b, \mathbf{h}_p\right)-{\operatorname{sim}}\left(\mathbf{h}_f, \mathbf{h}_p\right)+m\right),
\end{split}
\end{equation}
where $m$ represents a margin value, used to control the minimum gap between similarities.

\subsubsection{Counterfactual Inference Disentanglement}
In addition to contrastive disentanglement, we introduce a counterfactual inference strategy to enhance the disentanglement of focus and background information. Counterfactual inference is a process of exploring the consequences of alternative actions in hypothetical situations, which are usually contrary to the actual observed results under specific conditions \cite{mb20}. 

In this paper, the idea behind counterfactual inference is that, when either background or focus information is missing, the recommendation system's decision will rely on the remaining information source. This provides an opportunity to precisely identify and disentangle focus and background information. Specifically, for a dialogue  $\mathcal{C}_{k}$, we first determine whether the user's choice in the dialogue is primarily influenced by focus information or background information. To achieve this, we compute the similarity between the focal information and background information with the ground truth item respectively, thus generating pseudo labels for each conversation leader. Then Based on the idea of counterfactual inference, the similarity between the dominant factor and the target item is expected to be significantly greater than the similarity between the other factor and the target item. By identifying the dominant factor influencing the user's choice, this process facilitates the disentanglement of focus and background information.
\begin{equation}
\mathcal{L}_{c i}=\left\{\begin{array}{ll}
-\frac{\operatorname{sim}\left(\mathbf{h}_{f}, \mathbf{h}_{t}\right)}{\operatorname{sim}\left(\mathbf{h}_{f}, \mathbf{h}_{t}\right)+\operatorname{sim}\left(\mathbf{h}_{b}, \mathbf{h}_{t}\right)}, & \mathbf{h}_{t} \in D_f
\\
-\frac{\operatorname{sim}\left(\mathbf{h}_{b}, \mathbf{h}_{t}\right)}{\operatorname{sim}\left(\mathbf{h}_{b}, \mathbf{h}_{t}\right)+\operatorname{sim} \left(\mathbf{h}_{f}, \mathbf{h}_{t}\right)}, & \mathbf{h}_{t} \notin D_{f}
\end{array}\right.
\end{equation}
where $\mathbf{h}_{t}$ is the ground truth item. If $\mathbf{h}_{t} \in D_f$, then the target item is mainly dominated by the focus information, otherwise it means that the target item is dominated by the background information.

\subsection{Adaptive Prompt Learning Module}
After successfully disentangling the focus information and background information from the dialogue context, we design an adaptive prompt learning module to efficiently utilize these two types of information for accurate recommendations. This module first constructs a prompt pool and then uses a prompt selector to dynamically select the appropriate prompt from the prompt pool based on the current conversation.
\subsubsection{Prompt Pool}
To provide a variety of options, we construct the initial prompt pool $\mathbf{T}_{pool}$, which is defined as follows:
\begin{equation}
\mathbf{T}_{pool}=\left[\mathbf{T}^{1} ; \ldots; \mathbf{T}^{i} ; \ldots; \mathbf{T}^{\eta}\right],
\end{equation} 
\begin{equation}
\mathbf{T}_{f}^{i} = \left[\mathbf{h}_{f};\mathbf{h}_{cls}\right],
\end{equation} 
\begin{equation}
\mathbf{T}_{b}^{i} = \left[\mathbf{h}_{b};\mathbf{h}_{cls}\right],
\end{equation} 
where $\mathbf{T}^{i}$ consists of two components: $\mathbf{T}_{f}^{i}$ and $\mathbf{T}_{b}^{i}$. Then $\mathbf{T}_{f}^{i}$ and $\mathbf{T}_{b}^{i}$ respectively represent the fusion of the focus information $\mathbf{h}_{f}$ and background information $\mathbf{h}_{b}$ with the dialogue context $\mathbf{h}_{cls}$. 

\subsubsection{Prompt Selector}
After constructing the prompt pool, we design a prompt selector to choose the most pertinent prompt from the prompt pool, on the basis of the ongoing dialogue context. The overall process is as follows:
\begin{equation}
\mathbf{T}_{c}^{i} = \mathbf{T}_{f}^{i}\mathbf{W}_{f}^{i}+\mathbf{T}_{b}^{i} \mathbf{W}_{b}^{i},
\end{equation}
\begin{equation}
\mathbf{T}_{c}=\left[\mathbf{T}_{c}^{1}; \ldots; \mathbf{T}_{c}^{i} ; \ldots; \mathbf{T}_{c}^{\eta}\right],
\end{equation} 
\begin{equation}
\mathbf{T}_{select}=\mathbf{T}_{c}\left[\mathrm{F}_{select}(\mathbf{T}_{c})\right],
\end{equation}
where $\mathbf{W}_{f}^{i}$ and $\mathbf{W}_{b}^{i}$ controls the importance of the focus information $\mathbf{T}_{f}^{i}$ and the background information $\mathbf{T}_{b}^{i}$. And these weights is predefined and satisfy the condition $\mathbf{W}_{b}^{i}+\mathbf{W}_{f}^{i}=1$,  $\mathbf{W}_{b}^i, \mathbf{W}_{f}^i \in [0,1]$. $\mathbf{T}_{c}$ refers to the final prompt pool obtained after performing the weight-based fusion of the focus information and background information from the initial prompt pool.
$\mathrm{F}_{select}$ is a learnable neural network, e.g., a multilayer perceptron or an attention network, to select the index of best prompt based on $\mathbf{T}_{c}$. In this work, for simplicity, we use a multilayer perceptron. $\mathbf{T}_{select}$ represents the embedding corresponding to the best index in the prompt pool. 

\subsection{Downstream Tasks}
We further design task-specific prompts tailored for downstream recommendation and conversation tasks, which fully leverage the advantages of prompt learning to enhance the model's overall performance.

\subsubsection{Item Recommendation Task}
The recommendation task focuses on predicting items that align with user interests. Next, we introduce the item recommendation prompt and parameter optimization for the recommendation task.

\textbf{Item Recommendation Prompt.} For accurately capturing user interests in item recommendations, we construct a recommendation-oriented prompt. 
Specifically, the recommendation-oriented prompt consists of the selected context representation $\mathbf{T}_{select}$, recommendation-specific soft tokens $\mathbf{T}_{rec}^{init}$, and the original dialogue history $\mathcal{C}$, formally represented as follows:
\begin{equation}
\mathbf{T}_{rec}^{select}=\left[\mathbf{T}_{select}; \mathbf{T}_{rec}^{init}; \mathcal{C}\right],
\end{equation}
\begin{equation}
\mathbf{T}_{rec}^{kg}=\left[\mathbf{h}_\mathcal{G}; \mathbf{T}_{rec}^{init}; \mathcal{C}\right].
\end{equation}

Then, $\mathbf{T}_{rec}^{select}$ and $\mathbf{T}_{rec}^{kg}$ is separately fed into a frozen DialoGPT to produce the vector representation $\mathbf{h}_{rec}^{select}$ and $\mathbf{h}_{rec}^{kg}$, respectively. Finally, we adaptively fuse them into the final user representation $\mathbf{h}_{rec}$.
\begin{equation}
\mathbf{h}_{rec} =\mathbf{W}_{s} \mathbf{h}_{rec}^{select} + \mathbf{W}_{kg}\mathbf{h}_{rec}^{kg},
\end{equation}
where $\mathbf{W}_{s}$ and $\mathbf{W}_{kg}$ are learnable weight matrices.

\textbf{Parameter Optimization for Recommendation.} Following the training approach in \citet{mb35}, we first pre-train the item recommendation module to bridge the gap between prompts and contextual information. Formally, the prediction probability of an entity is formulated as follows:
\begin{equation}
P_{entity}=\operatorname{Softmax}\left(\mathbf{h}_{rec}^{\mathrm{T}} \mathbf{E}_{entity }\right),
    \label{eq:softmax}
\end{equation}
where $\mathbf{E}_{entity} \in \mathbb{R}^{d_e  \times N_{entity}}$ is the embedding matrix of all entities. In the pre-training stage, we employ cross-entropy as the loss function to optimize the model parameters:
\begin{equation}
\begin{split}
    \mathcal{L}_{pre} = & - \sum_{i=1}^{N_{(c)}} \sum_{e=1}^{N_{(e)}}  \left[ y_{i,e} \log(P_{entity}^{i,e}) +\right. \\
    & \left.  (1 - y_{i,e}) \log(1 - P_{entity}^{i,e}) \right]  +  \lambda(\mathcal{L}_{cd} + \mathcal{L}_{ci}),
\end{split}
\end{equation}
where $N_{(c)}$ represents the total number of conversation instances, $i$ denotes the index of a specific conversation, $N_{(e)}$ is the set of mentioned entities in the input context, and $e$ represents the index of an entity. After completing the pre-training, we also use the cross-entropy loss function to optimize the model parameters. The difference lies in the optimization targets: the pre-training stage focuses on entities, while the training stage focuses on items. The specific optimization formula is as follows:
\begin{equation}
\begin{split}
\mathcal{L}_{rec} =  - \sum_{i=1}^{N_{(c)}} \sum_{j=1}^M &\left[ y_{i,j} \log(P_{item}^{i,j}) +\right. \\
                     & \left.  (1 - y_{i,j}) \log(1 - P_{item}^{i,j}) \right],
\end{split}
\end{equation}
where $M$ denotes the total number of items, and $j$ serves as the index for an individual item. The variable $y_{i, j}$  represents the ground-truth label, indicating that item $i$ is the correct label for the $j$-th training instance. The probability $P_{item}^{i,j}$ of recommending candidate items is calculated with similar to eq. \eqref{eq:softmax}. In addition, during the training process, the parameters of the DialoGPT model are frozen.

\subsubsection{Response Generation Task}
In conversational recommendation, the purpose of the response generation task is to produce natural, relevant, and informative text responses to enhance user interaction experience. Next, we describe the prompt construction and parameter optimization for the response generation task. 

\textbf{Response Generation Prompt.} Similar to constructing prompts for recommendation tasks, we also design task-specific prompts for the response generation task. Specifically, we develop two types of prompts: a selection-enhanced prompt  $\mathbf{T}_{g e n}^{select}$ and a knowledge-enhanced prompt $\mathbf{T}_{g e n}^{kg}$. These are formalized as follows:
\begin{equation}
    \mathbf{T}_{g e n}^{select}=\left[\mathbf{T}_{select}; \mathbf{T}_{gen}^{init}; \mathcal{C}\right],
\end{equation}
\begin{equation}
    \mathbf{T}_{g e n}^{kg}=\left[\mathbf{h}_{\mathcal{G}}; \mathbf{T}_{gen}^{init}; \mathcal{C}\right],
\end{equation}
where $\mathbf{T}_{gen}^{init}$ represents the initial conversation-specific soft tokens, $\mathcal{C}$ is represented as a sequence of word tokens, which denotes the original conversation history. 

\textbf{Parameter Optimization for Response Generation.} After constructing the two prompts, we optimize the parameters $\Theta_{g e n}$ of the generation model by minimizing the specified objective function:
\begin{equation}
\begin{split}
    &\mathcal{L}_{gen} = \\& -\frac{1}{N_{(c)}} \sum_{i=1}^{N_{(c)}} \sum_{j=1}^{L_{gen}}  \log P_{gen}\left(w_{i, j} \mid \mathbf{T}_{gen}^{select}, \mathbf{T}_{gen}^{kg}; \Theta_{gen} ; w_{<j}\right),
\end{split}
\end{equation}
where $L_{gen}$ denotes the length of the generated response. $w_{< j}$ represents the sequence of words that appear before the $j$-th position. $P_{gen}(\cdot)$ indicates the probability of response generation. 
\section{EXPERIMENTS}              
To comprehensively demonstrate the effectiveness of our proposed DisenCRS model, we design and conduct extensive experiments to answer the following key questions: (1) \textbf{RQ1}: How does the performance of DisenCRS compare to existing baselines? (2) \textbf{RQ2}: What is the impact of the proposed dual contextual disentanglement module? (3) \textbf{RQ3}: Is our adaptive prompt learning module effective? (4) \textbf{RQ4}: How do different hyperparameters affect the performance of DisenCRS?

\subsection{Experimental Setup}
\subsubsection{Datasets.}
We conduct extensive experiments on two widely used conversational recommendation datasets, i.e., ReDial \cite{mb28} and INSPIRED \cite{mb37}, similar to previous work \cite{mb35,mb27}. The detailed statistics of these datasets are shown in Table \ref{tab:datas}. Both ReDial and INSPIRED are conversational recommendation datasets specifically designed for movie recommendations, aiming to simulate the interaction process between users and recommendation systems. 

\begin{table}[]
\caption{Statistics of datasets}
\label{tab:datas}
\centering
\begin{tabular}{l|c|c}
\hline
\textbf{Statistic} & \textbf{ReDial} & \textbf{INSPIRED} \\ \hline
\# users           & 956             & 1,482              \\
\# conversation    & 10,006           & 1,001              \\
\# utterance       & 182,150          & 35,811             \\
\# entities        & 64,364           & 17,321             \\ \hline
\end{tabular}
\end{table}

\begin{table*}[htbp]
\caption{Overall performance on recommendation task. $^*$ indicates a statistically significant difference with a p-value $<$ 0.05 compared to other baseline methods.  Compared with the baselines, our proposed model achieves the highest results on all metrics.}
\centering
\resizebox{\textwidth}{!}
{
\begin{tabular}{ccccccccccccc}
\toprule
Dataset & Metric & Popularity & TextCNN & BERT  & ReDial & KBRD & KGSF & TREA & VRICR & UniCRS & DCRS & \textbf{DisenCRS} \\ \midrule
\multirow{7}{*}{ReDial}   
    & Recall@1  & 0.011 & 0.010 & 0.027 & 0.010 & 0.033 & 0.035 & 0.045 & 0.054 & 0.065 & 0.076 & \textbf{0.081$^*$} \\
    & Recall@10 & 0.053 & 0.066 & 0.142 & 0.065 & 0.150 & 0.175 & 0.204 &  0.244 & 0.241 & 0.253 & \textbf{0.268$^*$} \\
    & Recall@50 & 0.183 &0.187 & 0.307 & 0.182 & 0.311 & 0.367 & 0.403 & 0.406 & 0.423 & 0.439 & \textbf{0.451$^*$} \\
    & NDCG@10 &0.029 & 0.033 &0.075  & 0.034 & 0.083 & 0.094 & 0.114 & 0.138 & 0.143 & 0.154 & \textbf{0.162$^*$} \\
    & NDCG@50 & 0.057 & 0.059  & 0.112  & 0.059 & 0.118 & 0.137 & 0.158 & 0.174 & 0.183 & 0.195 & \textbf{0.210$^*$} \\
    & MRR@10 & 0.021 &0.023 & 0.055   & 0.024 & 0.062 & 0.070 & 0.087 & 0.106 & 0.113 & 0.123 & \textbf{0.130$^*$} \\
    & MRR@50 & 0.027 & 0.028 &0.063    & 0.029 & 0.070 & 0.079 & 0.096 & 0.114 & 0.121 & 0.132 & \textbf{0.138$^*$} \\ \midrule
\multirow{7}{*}{INSPIRED} 
    & Recall@1 & 0.031 & 0.025 & 0.049  & 0.009 & 0.042 & 0.051 & 0.047 & 0.043 & 0.085 & 0.093 & \textbf{0.094$^*$} \\
    & Recall@10 & 0.155 & 0.119 & 0.189 & 0.048 & 0.135 & 0.132 & 0.146 & 0.141 & 0.230 & 0.226 & \textbf{0.252$^*$} \\
    & Recall@50 & 0.322 & 0.245 & 0.322 & 0.213 & 0.236 & 0.239 & 0.347 & 0.336 & 0.398 & 0.414 & \textbf{0.423$^*$} \\
    & NDCG@10 &0.085 & 0.066 & 0.112   & 0.023 & 0.088 & 0.092 & 0.095 & 0.091 & 0.149 & 0.153 & \textbf{0.165$^*$} \\
    & NDCG@50 &0.122 & 0.094 &0.141  & 0.059 & 0.109 & 0.114 & 0.132 & 0.134 & 0.187 & 0.192 & \textbf{0.200$^*$} \\
    & MRR@10 &0.064 &0.050 &0.088   & 0.015 & 0.073 & 0.079 & 0.076 & 0.075 & 0.125 & 0.130 & \textbf{0.139$^*$} \\
    & MRR@50 &0.071 &0.056 &0.095    & 0.023 & 0.077 & 0.083 & 0.087 & 0.085 & 0.133 & 0.137 & \textbf{0.146$^*$} \\ 
\bottomrule
\end{tabular}
}
\label{tab:performance}
\end{table*}

\subsubsection{Evaluation Metrics.}
Following previous work \cite{mb35,mb27},  we employ multiple evaluation metrics for the recommendation task, including Recall@k (k = 1, 10, 50), NDCG@k (k = 10, 50), and MRR@k (k = 10, 50). For the conversational task, both automatic and human evaluations are conducted. In the automatic evaluation, we report metrics such as BLEU-N (N = 2, 3), ROUGE-N (N = 2, L), and Distinct-N (N = 2, 3, 4). For the human evaluation, we randomly sample 20 responses and invite two annotators to score them. The scoring ranges from 1 to 3 and primarily focuses on two key aspects: informativeness and fluency.

\subsubsection{Baseline Methods.} We select the following competitive baselines for performance comparison, based on \citet{mb35}.

    (1) \textbf{Popularity}: It ranks the items based on their historical recommendation frequency. (2) \textbf{TextCNN} \cite{mb39}: It applies convolutional neural networks to extract contextual features for ranking items. (3) \textbf{BERT} \cite{mb40}: It is a pre-trained deep bidirectional transformer model that aims to improve natural language understanding through masked language modeling and next-sentence prediction tasks. (4) \textbf{GPT-2} \cite{mb41}: It serves as a fundamental baseline for text generation, benefiting from large-scale pre-trained language models. (5) \textbf{DialoGPT} \cite{mb42}: It is a large-scale generative pre-trained model optimized for conversational response generation, which leverages extensive dialogue data to produce contextually relevant and coherent responses. (6) \textbf{BART} \cite{mb43}: It aims to improve natural language generation and understanding tasks by reconstructing corrupted text. (7) \textbf{ReDial} \cite{mb28}: It proposes a dialogue module based on HRED \cite{mb48} and a recommendation module based on an autoencoder. (8) \textbf{KBRD} \cite{mb24}: It brings knowledge graphs to conversational recommendation, connecting recommendation and dialogue systems through knowledge propagation. (9) \textbf{KGSF} \cite{mb26}: It integrates word-oriented and entity-oriented knowledge graphs to enhance data representation in conversational recommendation.
    (10) \textbf{UniCRS} \cite{mb27}: It pioneers the unification of recommendation and dialogue tasks within the prompt learning paradigm and leverages pre-trained language models to accomplish both subtasks. (11)\textbf{TREA} \cite{mb29}: It introduces a multi-hierarchical tree structure for reasoning in CRSs, enhancing causal understanding of entities and improving response generation with historical context. (12) \textbf{VRICR} \cite{mb31}: It focuses on enhancing incomplete knowledge graphs using large-scale dialogue corpora and perform dynamic knowledge reasoning based on the dialogue context. (13) \textbf{DCRS}\cite{mb35}: It proposes a demonstration-augmented conversational recommendation system that improves the understanding of dialogue context by retrieving and learning demonstrations.

\subsubsection{Implementation Details}
In our experiments, we fine-tune the parameters of DisenCRS using the AdamW \cite{mb46}, with the learning rate set to 0.0005 for the recommendation task and 0.0001 for the dialogue task. The batch size is set to 24. 
We use the DialoGPT-small model, pre-trained on 147M Reddit dialogues, as the base language model and freeze all its parameters to ensure stability during training. The embedding dimension and the number of Transformer layers are 768 and 12, respectively. Additionally, for the recommendation subtask, we follow DCRS \cite{mb35} and use a fixed RoBERTa-base model to encode the input text semantically.

\subsection{Overall Performance (RQ1)}
In this section, we conduct extensive experiments and report the average metrics to  verify the effectiveness of our model in both recommendation and response generation tasks. 

\begin{table*}[]
\caption{Automatic evaluation results on the response generation task. Our model achieves the highest performance for the response generation task.} 
\label{tab:Automatic}
\scalebox{0.96}{
\begin{tabular}{c|ccccccc|ccccccl}
\hline
\multirow{3}{*}{\textbf{Model}} & \multicolumn{7}{c|}{\textbf{ReDial}}                                                                                                                                                                                                                                   & \multicolumn{7}{c}{\textbf{INSPIRED}}                                                                                                                                                                                                                               \\ \cline{2-15} 
                                & \multicolumn{2}{c|}{\textbf{BLEU}}                                        & \multicolumn{2}{c|}{\textbf{ROUGE}}                                       & \multicolumn{3}{c|}{\textbf{DIST}}                                                                              & \multicolumn{2}{c|}{\textbf{BLEU}}                                        & \multicolumn{2}{c|}{\textbf{ROUGE}}                                       & \multicolumn{3}{c}{\textbf{DIST}}                                                                           \\ \cline{2-15} 
                                & \multicolumn{1}{c|}{\textbf{-2}}    & \multicolumn{1}{c|}{\textbf{-3}}    & \multicolumn{1}{c|}{\textbf{-2}}    & \multicolumn{1}{c|}{\textbf{-L}}    & \multicolumn{1}{c|}{\textbf{-2}}    & \multicolumn{1}{c|}{\textbf{-3}}    & \textbf{-4}                         & \multicolumn{1}{c|}{\textbf{-2}}    & \multicolumn{1}{c|}{\textbf{-3}}    & \multicolumn{1}{c|}{\textbf{-2}}    & \multicolumn{1}{c|}{\textbf{-L}}    & \multicolumn{1}{c|}{\textbf{-2}}    & \multicolumn{1}{c|}{\textbf{-3}}    & \multicolumn{1}{c}{\textbf{-4}} \\ \hline
DialoGPT                       & \multicolumn{1}{c|}{0.041}          & \multicolumn{1}{c|}{0.021}          & \multicolumn{1}{c|}{0.054}          & \multicolumn{1}{c|}{0.258}          & \multicolumn{1}{c|}{0.436}          & \multicolumn{1}{c|}{0.632}          & 0.771                               & \multicolumn{1}{c|}{0.031}          & \multicolumn{1}{c|}{0.014}          & \multicolumn{1}{c|}{0.041}          & \multicolumn{1}{c|}{0.207}          & \multicolumn{1}{c|}{1.954}          & \multicolumn{1}{c|}{2.750}          & 3.235                           \\
GPT-2                            & \multicolumn{1}{c|}{0.031}          & \multicolumn{1}{c|}{0.013}          & \multicolumn{1}{c|}{0.041}          & \multicolumn{1}{c|}{0.244}          & \multicolumn{1}{c|}{0.405}          & \multicolumn{1}{c|}{0.603}          & 0.757                               & \multicolumn{1}{c|}{0.026}          & \multicolumn{1}{c|}{0.011}          & \multicolumn{1}{c|}{0.034}          & \multicolumn{1}{c|}{0.212}          & \multicolumn{1}{c|}{2.119}          & \multicolumn{1}{c|}{3.084}          & 3.643                           \\
BART                            & \multicolumn{1}{c|}{0.024}          & \multicolumn{1}{c|}{0.011}          & \multicolumn{1}{c|}{0.031}          & \multicolumn{1}{c|}{0.229}          & \multicolumn{1}{c|}{0.432}          & \multicolumn{1}{c|}{0.615}          & 0.705                               & \multicolumn{1}{c|}{0.018}          & \multicolumn{1}{c|}{0.008}          & \multicolumn{1}{c|}{0.025}          & \multicolumn{1}{c|}{0.208}          & \multicolumn{1}{c|}{1.920}          & \multicolumn{1}{c|}{2.501}          & 2.670                           \\
ReDial                          & \multicolumn{1}{c|}{0.004}          & \multicolumn{1}{c|}{0.001}          & \multicolumn{1}{c|}{0.021}          & \multicolumn{1}{c|}{0.187}          & \multicolumn{1}{c|}{0.058}          & \multicolumn{1}{c|}{0.204}          & 0.442                               & \multicolumn{1}{c|}{0.001}          & \multicolumn{1}{c|}{0.000}          & \multicolumn{1}{c|}{0.004}          & \multicolumn{1}{c|}{0.168}          & \multicolumn{1}{c|}{0.359}          & \multicolumn{1}{c|}{1.043}          & 1.760                           \\
KBRD                            & \multicolumn{1}{c|}{0.038}          & \multicolumn{1}{c|}{0.018}          & \multicolumn{1}{c|}{0.047}          & \multicolumn{1}{c|}{0.237}          & \multicolumn{1}{c|}{0.070}          & \multicolumn{1}{c|}{0.288}          & 0.488                               & \multicolumn{1}{c|}{0.021}          & \multicolumn{1}{c|}{0.007}          & \multicolumn{1}{c|}{0.029}          & \multicolumn{1}{c|}{0.218}          & \multicolumn{1}{c|}{0.416}          & \multicolumn{1}{c|}{1.375}          & 2.320                           \\
KGSF                            & \multicolumn{1}{c|}{0.030}          & \multicolumn{1}{c|}{0.012}          & \multicolumn{1}{c|}{0.039}          & \multicolumn{1}{c|}{0.244}          & \multicolumn{1}{c|}{0.061}          & \multicolumn{1}{c|}{0.278}          & 0.515                               & \multicolumn{1}{c|}{0.023}          & \multicolumn{1}{c|}{0.007}          & \multicolumn{1}{c|}{0.031}          & \multicolumn{1}{c|}{0.228}          & \multicolumn{1}{c|}{0.418}          & \multicolumn{1}{c|}{1.496}          & 2.790                           \\
VRICR                           & \multicolumn{1}{c|}{0.021}          & \multicolumn{1}{c|}{0.008}          & \multicolumn{1}{c|}{0.027}          & \multicolumn{1}{c|}{0.137}          & \multicolumn{1}{c|}{0.107}          & \multicolumn{1}{c|}{0.286}          & 0.471                               & \multicolumn{1}{c|}{0.011}          & \multicolumn{1}{c|}{0.001}          & \multicolumn{1}{c|}{0.025}          & \multicolumn{1}{c|}{0.187}          & \multicolumn{1}{c|}{0.853}          & \multicolumn{1}{c|}{1.801}          & 2.827                           \\
TREA                            & \multicolumn{1}{c|}{0.022}          & \multicolumn{1}{c|}{0.008}          & \multicolumn{1}{c|}{0.039}          & \multicolumn{1}{c|}{0.175}          & \multicolumn{1}{c|}{0.242}          & \multicolumn{1}{c|}{0.615}          & 1.176                               & \multicolumn{1}{c|}{0.013}          & \multicolumn{1}{c|}{0.002}          & \multicolumn{1}{c|}{0.027}          & \multicolumn{1}{c|}{0.195}          & \multicolumn{1}{c|}{0.958}          & \multicolumn{1}{c|}{2.565}          & 3.411                           \\
UniCRS                          & \multicolumn{1}{l|}{0.045}          & \multicolumn{1}{l|}{0.021}          & \multicolumn{1}{l|}{0.058}          & \multicolumn{1}{l|}{0.285}          & \multicolumn{1}{l|}{0.433}          & \multicolumn{1}{l|}{0.748}          & \multicolumn{1}{l|}{1.003}          & \multicolumn{1}{l|}{0.022}          & \multicolumn{1}{l|}{0.009}          & \multicolumn{1}{l|}{0.029}          & \multicolumn{1}{l|}{0.212}          & \multicolumn{1}{l|}{2.686}          & \multicolumn{1}{l|}{4.343}          & 5.520                           \\
DCRS                            & \multicolumn{1}{l|}{0.048}          & \multicolumn{1}{l|}{0.024}          & \multicolumn{1}{l|}{0.063}          & \multicolumn{1}{l|}{0.285}          & \multicolumn{1}{l|}{0.779}          & \multicolumn{1}{l|}{1.173}          & \multicolumn{1}{l|}{1.386}          & \multicolumn{1}{l|}{0.033}          & \multicolumn{1}{l|}{0.014}          & \multicolumn{1}{l|}{0.045}          & \multicolumn{1}{l|}{0.229}          & \multicolumn{1}{l|}{3.950}          & \multicolumn{1}{l|}{5.729}          & 6.233                           \\ \hline
\textbf{DisenCRS}               & \multicolumn{1}{c|}{\textbf{0.050$^*$}} & \multicolumn{1}{l|}{\textbf{0.027$^*$}} & \multicolumn{1}{l|}{\textbf{0.065$^*$}} & \multicolumn{1}{l|}{\textbf{0.286$^*$}} & \multicolumn{1}{l|}{\textbf{0.784$^*$}} & \multicolumn{1}{l|}{\textbf{1.212$^*$}} & \multicolumn{1}{l|}{\textbf{1.411$^*$}} & \multicolumn{1}{c|}{\textbf{0.034$^*$}} & \multicolumn{1}{l|}{\textbf{0.017$^*$}} & \multicolumn{1}{l|}{\textbf{0.048$^*$}} & \multicolumn{1}{l|}{\textbf{0.232$^*$}} & \multicolumn{1}{l|}{\textbf{4.014$^*$}} & \multicolumn{1}{l|}{\textbf{6.318$^*$}} & \textbf{6.623$^*$}                  \\ \hline
\end{tabular}}
\end{table*}

\subsubsection{Evaluation on Recommendation Task}
Table \ref{tab:performance} shows the recommendation performance of various models on the ReDial and INSPIRED datasets in terms of Recall@k, NDCG@k, and MRR@k metrics. The experimental results demonstrate that our model DisenCRS significantly outperforms other models across all metrics. Analysis of this table reveals the following phenomena:

(1) Popularity and TextCNN, as traditional recommendation approaches, exhibit mediocre performance on both datasets, indicating that these methods struggle to effectively capture contextual information in conversational recommendation scenarios. BERT demonstrates superior performance compared to traditional methods, yet it remains inferior to most conversational recommendation models. This suggests that although BERT can understand the semantic information of conversational context, it still falls short in modeling user interests effectively.

(2) In conversational recommendation models, ReDial serves as a classic baseline. However, due to its limitations in entity and contextual modeling, its performance is suboptimal. KBRD is the first to introduce an external knowledge graph for entity modeling, resulting in improved recommendation performance. Subsequently, KGSF further integrates word and entity knowledge graphs based on KBRD, optimizing data representation and thus improving the model's overall performance. Compared with KBRD and KGSF, TREA improved results by better mining entity-related information, e.g., introducing tree-structure reasoning to uncover the causal relationships between entities and retrieving similar conversations through entity information.
VRICR further improved recommendation performance by reconstructing missing entity relationships in incomplete knowledge graphs. 
Notably, UniCRS and DCRS exhibit outstanding performance compared to other baselines. Both of them combine entity representation and semantic information of conversational contexts, and incorporate well-designed prompts into DialoGPT for downstream tasks. In comparison with UniCRS, DCRS achieves relatively better performance, mainly because it employs a dialogue retrieval approach, fully leveraging richer contextual information to enhance the accuracy of recommendations.

(3) Finally, DisenCRS achieves the best performance on both datasets compared to the aforementioned baseline models. This achievement is attributed to its effective extraction of focus and background information in the conversation through contrastive disentanglement and counterfactual inference disentanglement, as well as its dynamic utilization of this information through the adaptive prompt learning module. Unlike existing methods that process the entire conversational context, DisenCRS more accurately utilizes the key information in the context, avoids the interference of minor information, and thus improves the accuracy of recommendations.

\subsubsection{Evaluation on Response Generation Task}
In this section, to comprehensively and accurately evaluate the effectiveness of the proposed model in the response generation task, we employ two evaluation methods: automatic evaluation and human evaluation.

\textbf{Automatic Evaluation.} 
In Table \ref{tab:Automatic}, DialoGPT, GPT-2, and BART demonstrate superior performance for response generation over most conversational recommendation models like ReDial, VRICR, and TREA, mainly because they are pre-trained on large-scale corpora and are able to generate informative and semantically consistent responses in CRS. UniCRS and DCRS are further optimized based on the DialoGPT model, thereby exhibiting superior performance. In contrast, DisenCRS achieves the best scores on BLEU, ROUGE, and DIST metrics, demonstrating its overall advantages in dialogue generation. This may be due to its effective disentanglement of contextual information, which allows the model to focus on relevant aspects of the dialogue while minimizing irrelevant noise, thereby producing more coherent and contextually appropriate responses.

\begin{table}[t] 
\centering \caption{Human evaluation results on the response generation task.} 
\label{tab:Human_Evaluation} 
\begin{tabular}{ccc} \hline \textbf{Model} & \textbf{Fluency} & \textbf{Informativeness} \\ \hline KBRD & 2.32 & 1.97 \\ KGSF & 2.46 & 2.05 \\ VRICR & 2.37 & 2.17 \\ UniCRS & 2.71 & 2.53 \\ DCRS & 2.80 & 2.65 \\ \hline \textbf{DisenCRS} & \textbf{2.85$^*$} & \textbf{2.72$^*$} \\ \hline \end{tabular} 
\end{table}

\textbf{Human Evaluation.} To further validate the effectiveness of the DisenCRS model on the response generation task, we conduct a human evaluation, 
with the results shown in Table \ref{tab:Human_Evaluation}. 
As observed from Table \ref{tab:Human_Evaluation}, DisenCRS outperforms other baselines in terms of the fluency and informativeness of the generated responses, demonstrating its advantage in producing coherent and information-rich responses. We believe that this advantage stems from the model's effective decoupling of focus and background information based on the dialogue history, which avoids the interference of extraneous information, thereby generating responses that are more contextually consistent and content-rich. 

\subsection{Impact of Dual Contextual Disentanglement Module (RQ2)}
Here, to demonstrate the effectiveness of the proposed disentanglement method, we conduct ablation experiments. Specifically, we design the following three variants: - w/o CD, which refers to the removal of contrastive disentanglement; - w/o CID, which eliminates the counterfactual inference disentanglement; and - w/o Dual, which indicates the absence of the contextual disentanglement module.
\begin{figure}[t]
	\begin{center}
		\includegraphics[width=1\linewidth]{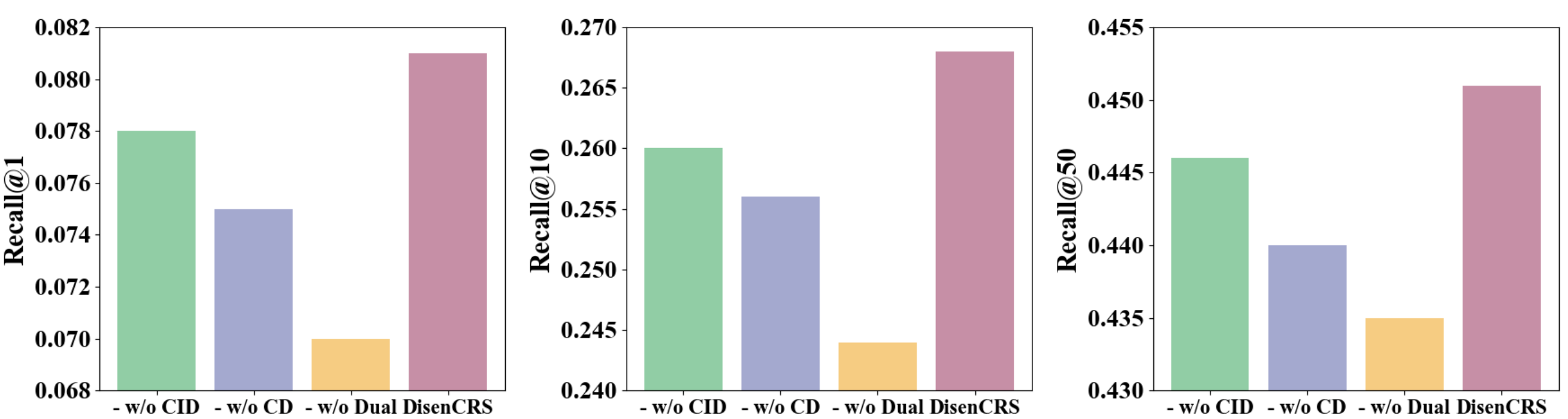}
	\end{center}
    \caption{Ablation comparison of the dual contextual disentanglement module.}
	\label{fig: distanglement}
\end{figure}
The performance of the aforementioned variants on the ReDial dataset is illustrated in Figure \ref{fig: distanglement}. Upon examination, we can draw the following conclusions: (1) DisenCRS achieves better results than - w/o CD, which indicates that the contrastive disentanglement mechanism effectively helps the model distinguish between focus information and background information, thereby better understanding the semantic structure of contextual conversations. (2) DisenCRS outperforms - w/o CID, which shows that by introducing a counterfactual reasoning mechanism, the model can better simulate the information distribution under different contextual assumptions, thereby enhancing the ability to capture user intentions. (3) - w/o Dual has the worst effect, which further illustrates the effectiveness of the above two disentanglement strategies.

\subsection{Effectivenss of Adaptive Prompt Learning Module (RQ3)}
To comprehensively evaluate the effectiveness of our adaptive prompt learning module, we conduct comparative experiments. We define the following variants: (1) DisenCRS-fc: represents using only the focus information from the dialogue context. (2) DisenCRS-bg: represents using only the background information from the dialogue context. (3) DisenCRS-fw: applies weighted fusion by manually setting fixed weights. We perform a grid search to select the best manual weights for the experiments.

The experimental results of the aforementioned variants are shown in Table \ref{tab:prompt_selector}. Both DisenCRS-fc and DisenCRS-bg perform worse than the full DisenCRS model, indicating that neither the background information nor the focus information can be discarded. They both play an important role in understanding user intentions. And DisenCRS-fc outperforms DisenCRS-bg, indicating that in most dialogues, focus information better reflects the user's true interests than background information. Furthermore, the DisenCRS-fw fusion strategy performs worse than the adaptive prompt learning module, suggesting that simply relying on fixed weights to fuse focus information and background information in dialogue makes it difficult to cope with diverse contextual requirements. They fail to fully leverage the complementary relationship between focus and background information. Overall, the adaptive prompt learning module is capable of dynamically selecting focus and background information based on the dialogue context, effectively filtering out noise while preserving key useful information. 

\begin{table}[]
\caption{Comparison of the adaptive prompt learning module with other methods.}
\label{tab:prompt_selector}
\begin{tabular}{ccccc}
\hline
\textbf{Method}   & \textbf{Recall@10} & \textbf{Recall@50} & \textbf{DIST@3} & \textbf{DIST@4} \\ \hline
DisenCRS-fc       & 0.248              & 0.430              & 0.743           & 1.164           \\
DisenCRS-bg       & 0.234              & 0.425              & 0.727           & 1.088           \\
DisenCRS-fw       & 0.259              & 0.442              & 0.967           & 1.302           \\
\hline
\textbf{DisenCRS} & \textbf{0.268$^*$}     & \textbf{0.451$^*$}     & \textbf{1.212$^*$}  & \textbf{1.411$^*$}  \\ \hline
\end{tabular}
\end{table}

\subsection{Hyperparameter Study (RQ4)}
\subsubsection{The impact of the number of prompts $\eta$ in the prompt pool}

\begin{figure}[t]
	\begin{center}
		\includegraphics[width=1\linewidth]{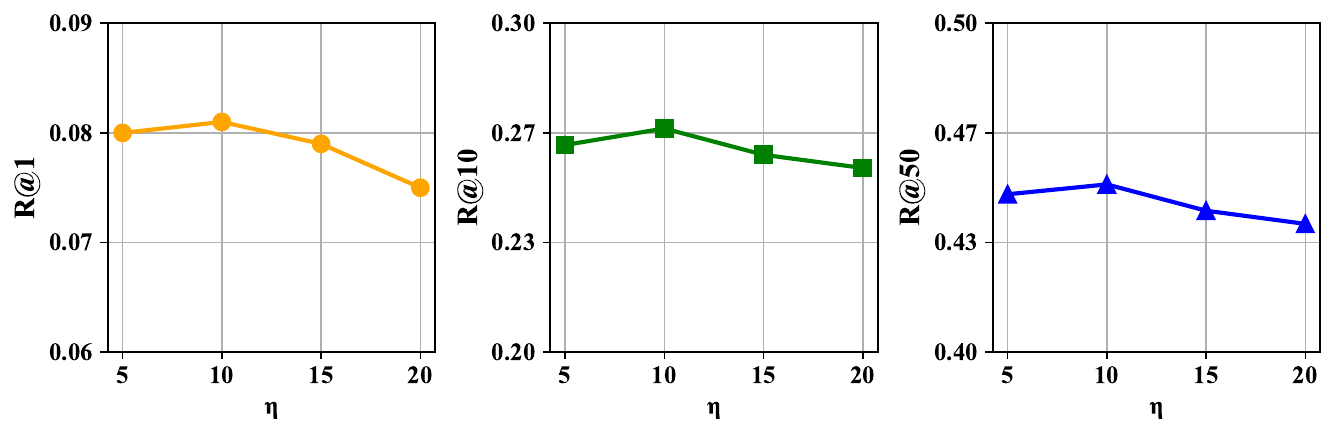}
	\end{center}
	\caption{Comparison of the number of prompts $\eta$ in the prompt pool.}
	\label{fig: prompts}
\end{figure}

To investigate the impact of the number of prompts in the prompt pool on model performance, we set different numbers of prompts $\eta$ (such as 5, 10, 15, and 20), as shown in Figure \ref{fig: prompts}. When the number of prompts in the prompt pool is small, the weight fusion granularity of the focus information and background information is coarse, making the model difficult to find the most suitable weight combination in different dialogue context scenarios. As the number of prompts in the prompt pool increases, the granularity of weight distribution becomes finer, enabling the model to select more appropriate weights for different dialogue scenarios, thereby improving performance. When the number of prompts in the prompt pool becomes excessive (e.g., $\eta$>10), the differences between weights may be too subtle, placing a higher challenge for the prompt selector and leading to performance degradation. 

\subsubsection{The impact of contextual disentanglement loss weights $\lambda$} 
\begin{figure}[t]
	\begin{center}
\includegraphics[width=1\linewidth]{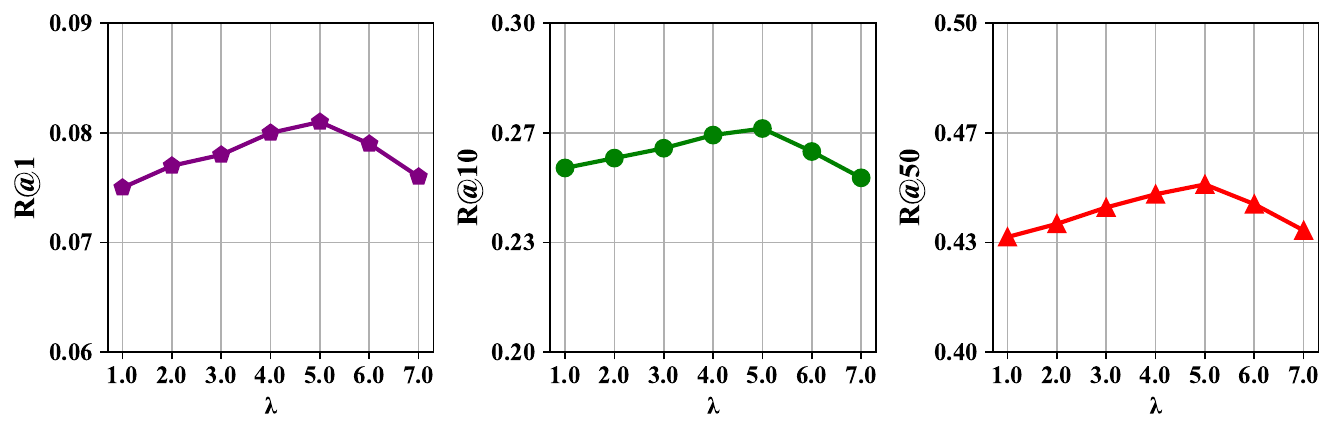}
	\end{center}
	\caption{Comparison of the contextual disentanglement loss weights $\lambda$.}
	\label{fig: weight}
\end{figure}

To investigate the impact of the contextual disentanglement loss on model performance, we introduce a weighting factor $\lambda$ before the disentanglement loss. By adjusting this weighting factor (ranging from 1.0 to 7.0), we conduct multiple experiments to analyze its effect on the recommendation task. As shown in Figure \ref{fig: weight}, experimental results show that the weight of the contextual disentanglement loss has a significant impact on the overall performance of the model. When the weight is too small, the disentanglement effect is insufficient, negatively affecting the recommendation performance; when the weight is too large, excessive disentanglement weakens the performance of the recommendation task. 
\section{CONCLUSION AND FUTURE WORK}
In this paper, we propose a contextual disentanglement approach for conversational recommendation, named DisenCRS. DisenCRS integrates contrastive disentanglement and counterfactual inference disentanglement techniques to effectively separate the focus information and the background information in the conversation context in an unsupervised manner. Moreover, we design an adaptive prompt learning module that can adaptively perceive the context information and dynamically select a suitable prompt from the prompt pool to support downstream recommendation and response generation tasks. Our extensive experiments on two benchmarking datasets verify the effectiveness of our DisenCRS model.

Although this paper introduces and highlights the potential of contextual disentanglement for advancing research on CRS, it represents a preliminary exploration and there is still room for improvement. In the future, we plan to explore large language models for disentanglement to enhance CRS, improving their ability to recognize different types of information in context. 
\begin{acks}
This research was supported by the National Natural Science Foundation of China (62402093) and, the Sichuan Science and Technology Program (2025ZNSFSC0479). This work was also supported in part by the National Natural Science Foundation of China under grants U20B2063 and 62220106008, and the Sichuan Science and Technology Program under Grant 2024NSFTD0034. 
\end{acks}

\clearpage



\bibliographystyle{ACM-Reference-Format}
\bibliography{sample-base}

\appendix

\end{document}